\documentclass[twocolumn,floatfix,showpacs,preprintnumbers,amsmath,amssymb,aps,prl]{revtex4}

\usepackage{amssymb,amsmath,graphics,epsfig}
\usepackage[usenames,dvipsnames]{color}

\begin{document}

\title{Laser radiation pressure slowing of a molecular beam}
\author{J.F. Barry*, E.S. Shuman, E.B. Norrgard, and D. DeMille}
\affiliation{Department of Physics, Yale University, P.O. Box 208120, New Haven, CT 06520, USA}

\date{\today}

\maketitle

\newpage
\cleardoublepage
\newpage
\cleardoublepage

{\bf There is substantial interest in producing samples of ultracold molecules for possible applications in quantum computation \cite{DeMille2002,Andre2006}, quantum simulation of condensed matter systems \cite{Goral2002,Micheli2006}, precision measurements \cite{Hudson2011,Cahn2008}, controlled chemistry \cite{Balakrishnan2001,Krems2008}, and high precision spectroscopy \cite{Flambaum2007}. A crucial step to obtaining large samples of ultracold, trapped molecules is developing a means to bridge the gap between typical molecular source velocities ($\sim\!150\!-\!600\frac{m}{s}$) and velocities for which trap loading or confinement is possible ($\lesssim 5\!-\!20\frac{m}{s}$). Here we show deceleration of a beam of neutral strontium monofluoride (SrF) molecules using radiative force. Under certain conditions, the deceleration results in a substantial flux of molecules with velocities $\lesssim\!\!50 \frac{m}{s}$. The observed slowing, from $\sim\!\!140\frac{m}{s}$, corresponds to scattering $\gtrsim\!10^4$ photons. We also observe longitudinal velocity compression under different conditions. Combined with molecular laser cooling techniques \cite{Stuhl2008,Shuman2009,Shuman2010}, this lays the groundwork to create slow and cold molecular beams suitable for trap loading.}

Tremendous advances have been made in the deceleration of molecular beams in the past decade.  Stark deceleration \cite{Bethlem99}, Zeeman deceleration \cite{Narevicius2008b}, and collisional deceleration \cite{Chandler2003} have all been demonstrated to slow molecular beams. However, only for fairly light species ($\sim\!20$ amu) with substantial vapor pressure at room temperature have these methods been demonstrated to allow slowing to velocities as low as a few m/s, as needed to make trapping possible \cite{Krems2009Meijer}. Optical deceleration has been demonstrated to slow molecular beams to rest \cite{Bishop2010}, but the high laser intensities required limit application to small volumes. While these methods are useful, all of them conserve phase-space density and hence slow without cooling.  However, recently it has been demonstrated that radiative forces can be used to cool molecules \cite{Shuman2010}. Assuming a given species is amenable to laser cooling, the same radiative forces can be used for slowing. As is well known from atoms \cite{Phillips1982,Prodan1982}, laser slowing can be effective over broad velocity ranges and is insensitive to position, so that it can work on a large phase-space volume of molecules. Laser slowing can also lead to simultaneous longitudinal velocity compression, which is advantageous for loading traps. Once trapped, these molecules may be further laser-cooled to increase the phase-space density. Finally, laser slowing methods appear viable for a variety of species, including some for which current phase-space-conserving slowing methods appear ill-suited; this is useful for addressing the needs of the variety of applications envisioned for ultracold molecules \cite{Carr2009}.

\begin{figure}
\includegraphics[height=4.0in]
{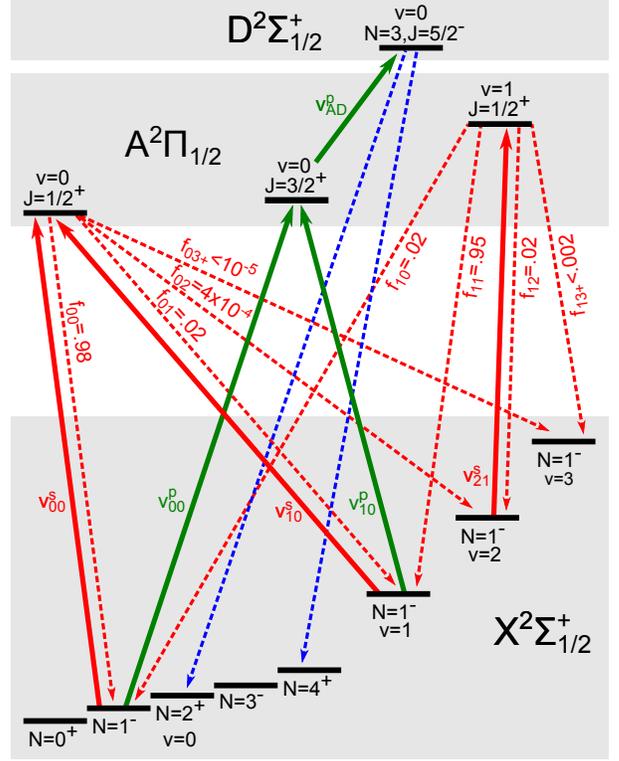}
\caption{{\bf Relevant energy levels and transitions in SrF.} Solid red lines denote slowing lasers, while solid green lines denote probe lasers. Dashed lines denote spontaneous decay channels, and FCFs ($f_{v'v}$) are labeled if applicable. Blue dashed lines indicate decay fluorescence channels used to determine the beam Doppler profile.} \label{fig:energylevelsanddetection}
\end{figure}

\begin{figure}
\includegraphics[height=1.3in]
{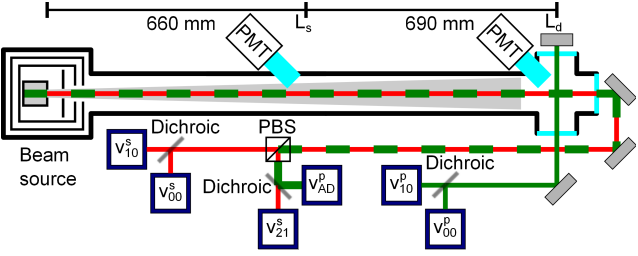}
\caption{{\bf Schematic of the apparatus.} Red (green) lines indicate slowing (probe) laser beams.} \label{fig:experimentalslowingdiagramnew}
\end{figure}

\begin{figure}
\includegraphics[height=5.0in]
{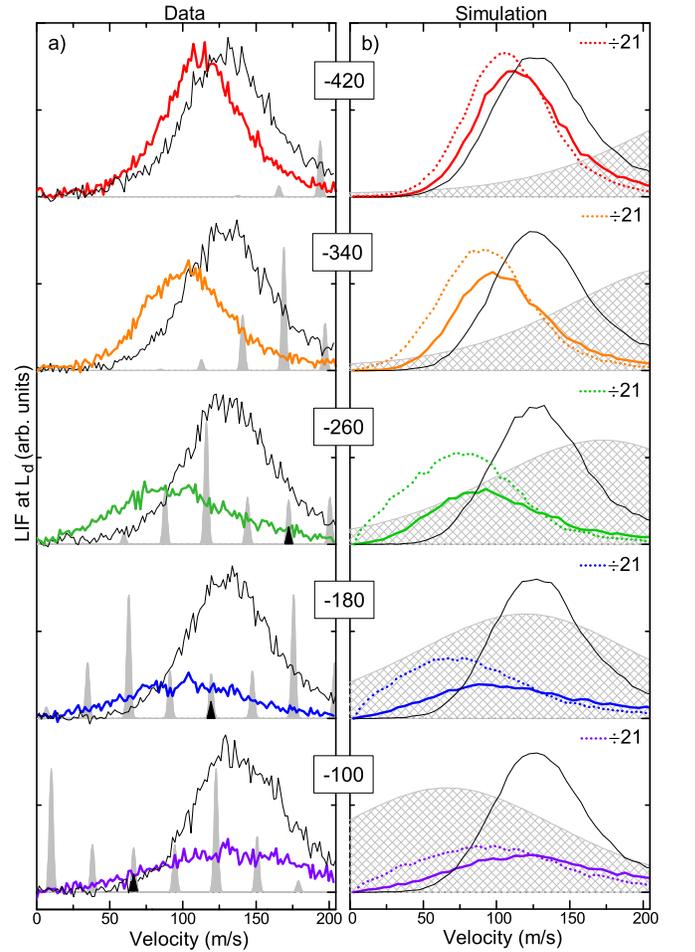}
\caption{{\bf Measured and simulated slowing for different detunings of the main slowing laser.} a) Measured slowed LVP (solid color), control LVP ({\bf\large \textendash}), and velocities corresponding to the $v_{00}^s$ rf sideband spectrum (gray, with center $\blacktriangle$). The panels are scaled so that all controls have equal heights. b) Simulated slowed LVP (solid color), simulated control LVP ({\bf\large \textendash}), and simulated slowed LVP with transverse cooling the entire length of the slowing (dotted color, scaled by 1/21). The gray shaded area indicates the assumed force versus velocity profile used in the simulation. The $\Delta v_{00}^s$ detuning (in MHz) is shown in the centered box for each panel set. The scaling of the simulated LVP with transverse cooling is chosen to accurately illustrate how the increased divergence of the slowest molecules impacts the apparent loss in our slowed LVP data.  In particular, the factor 21 is the ratio of the peaks of the simulated control LVPs with and without transverse cooling. This scaling hence factors out the simple gain in downstream flux due to transverse cooling; the different shapes of simulated slowed LVPs with and without transverse cooling show clearly the increased apparent loss of slow molecules in our experimental data, merely due to their increased divergence (since transverse cooling is not applied in these experiments) (Supplementary Discussion 1).
} \label{fig:slowinganddivergence}
\end{figure}

Here we experimentally demonstrate deceleration of a molecular beam using radiation pressure. This work builds upon similar results demonstrating deflection \cite{Shuman2009} and transverse cooling \cite{Shuman2010} of a molecular beam by radiative force. The crucial enabling feature for radiative slowing is the ability to scatter $\gtrsim\!10^4$ photons without heating the internal degrees of freedom of the molecules. Our scheme for creating this quasi-cycling transition \cite{Shuman2009,Shuman2010} is recounted briefly here and depicted in Fig. \ref{fig:energylevelsanddetection}. We employ the X$^2\Sigma^+_{1/2}(v\!\!=\!\!0,N\!\!\!=\!\!\!1)\!\rightarrow\!A^2\Pi_{1/2}(v'\!\!\!=\!\!0,J'\!\!\!=\!\!\!1/2)$ electronic transition of SrF, with $\tau\!\!=\!24$ ns lifetime, for cycling and slowing. We denote by $v_{00}^s$ the main cycling and slowing laser, with wavelength $\lambda_{00}^s$ and detuning from resonance $\Delta v_{00}^s$. The favorable Franck-Condon factors (FCFs) of SrF limit vibrational branching \cite{DiRosa2004}. Separate repump lasers, denoted $v_{10}^s$ and $v_{21}^s$, address residual vibrational leakage. Driving an $N\!\!=\!\!1\!\!\rightarrow\!\!J'\!\!=\!\!1/2$ transition eliminates rotational branching \cite{Stuhl2008}, while a magnetic field remixes the resulting dark ground-state Zeeman sublevels \cite{Berkeland2002}. This scheme should allow $\gtrsim \!10^5$ photon scattering cycles before the bright state population is reduced by $1/e$ \cite{Shuman2009,Shuman2010}.

We use the Doppler shift of laser-induced fluorescence (LIF) to measure the longitudinal velocity profile (LVP) of the molecular beam, as shown in Fig. \ref{fig:experimentalslowingdiagramnew}. Our detection scheme employs two perpendicular probe lasers, denoted $v_{00}^p$ and $v_{10}^p$, to excite molecules from the X$(v\!\!=\!\!0,1;N\!\!=\!\!1)$ states (with resolved spin-rotation structure (SRS) and hyperfine structure (HFS)) to the A$(v\!\!=\!\!0,J\!\!=\!\!\frac{3}{2})$ state (unresolved HFS) at a distance $L_d\!\!=\!\!1350$ mm downstream from the source. A longitudinal probe laser, denoted $v_{AD}^p$, then excites the molecules to the D$(v\!\!=\!\!0,N\!\!=\!\!3,J\!\!=\!\!\frac{5}{2})$ state (unresolved HFS). Monitoring the D$\rightarrow$X LIF as a function of the $v_{AD}^{p}$ laser frequency yields a Doppler-shifted LVP free of SRS/HFS, at a wavelength easily filtered from all laser light.

The $v^s_{00}$, $v^s_{10}$, $v^s_{21}$, and $v^p_{AD}$ lasers are spatially overlapped counterpropagating to the molecular beam. To address all SRS/HFS levels over a wide velocity range, the $v^s_{00}$, $v^s_{10}$, and $v^s_{21}$ lasers have radio-frequency (rf) sidebands with modulation index $m\!=\!3.1$ and modulation frequency $41 \text{ MHz} <\!f_{mod}\! < 44 \text{ MHz}$, unless noted otherwise. Due to the large frequency extent of the sidebands, we do not expect longitudinal velocity compression \cite{Zhu1991}. We note that the dark magnetic sublevels of the X$(N=1)$ state prevent use of a Zeeman slower. The slowing lasers are not chirped \cite{Prodan1984} due to the temporal extent of the molecular beam pulse ($\sim\!\!10$ ms). Magnetic field coils create an approximately uniform field $B\!\!=\!\!9$ G at an angle $\theta_{B}\!=\!45^{\circ}$ relative to the $v^s_{00}$ linear polarization over the length $120 \text{ mm}\!\lesssim\!L\! \lesssim\! 1350$ mm. A supplementary light detector at $L_s\!\!=\!660$ mm allows monitoring of the LIF from spontaneously emitted photons during cycling.

Application of the slowing lasers shifts the molecular beam LVP, as shown in Fig. \ref{fig:slowinganddivergence}a for various $\Delta v_{00}^s$. As $\Delta v^{s}_{00}$ is tuned towards $\langle v_{f} \rangle / \lambda_{00}^s$ from the red (where $\langle v_{f} \rangle$ is the mean forward velocity), the LVP is shifted to lower velocities, until $\Delta v_{00}^s \!\!\approx\!\! \langle v_{f} \rangle / \lambda_{00}^s$; then, when tuned further to the blue, the LVP gradually returns to its unperturbed state. However, the shift to lower velocities is accompanied by an increase in apparent molecule loss, which is most severe when $\Delta v^{s}_ {00} \!\!\approx\!\!\langle v_{f} \rangle / \lambda_{00}^s$.

We argue that this apparent loss is due primarily to increased divergence and transverse heating as the beam is slowed. Several other loss mechanisms were ruled out as the dominant cause after investigation. Increasing the background gas pressure (primarily helium) by $5\times$ changed the slowed LVPs little, indicating that background gas collisions are not a dominant loss mechanism. We investigated possible loss to other rovibrational states which could arise from various mechanisms. For example, off-resonant excitation to the A$(v\!\!=\!\!0,J\!\!=\!\!\frac{3}{2})$ state by the $v_{00}^s$ laser or HFS mixing in the A$(v\!\!=\!\!0,J\!\!=\!\!\frac{1}{2})$ state could transfer population to the dark X$(v\!\!=\!\!0,N\!\!=\!\!3)$ state; stray electric fields could lead to decays from the A$(v\!\!=\!\!0,J\!\!=\!\!\frac{1}{2})$ state to the dark X$(v\!\!=\!\!0;N\!\!=\!\!0,2)$ states; or the $v_{AD}^p$ laser could off-resonantly excite molecules from the A$(v\!\!=\!\!0,J\!\!=\!\!\frac{1}{2})$ state to the D$(v\!\!=\!\!0,N\!\!=\!\!1)$ state before they reach $L_d$. To investigate such mechanisms, we explicitly probed the populations of the X$(v\!\!=\!\!0;N\!\!=\!\!0,2,3)$ and X$(v\!\!=\!\!1,N\!\!=\!\!0)$ states and determined that $<\!\!10\%$ combined total loss could be attributed to such processes. Loss to the X$(v\!\!=\!\!3,N\!\!=\!\!1)$ state was not directly measured, but was estimated from the observed increase in spontaneous scattering LIF at $L_s$ by adding the $v_{21}^s$ repump laser; this indicated that molecules cycled through the X$(v\!\!=\!\!2,N\!\!=\!\!1)$ state $\sim\!3\!\times$ before reaching $L_d$. Together with the estimated FCFs \cite{Shuman2009}, this yields an estimated $\sim\!6\%$ loss to the X$(v\!\!=\!\!3,N\!\!=\!\!1)$ state. Overall, our inability to find evidence of population of dark states, together with the observed LVP shift of $\sim\!50$ m/s, is preliminary evidence that our quasi-cycling transition is nearly closed for up to $\sim\!\!10^4$ scattered photons. Additional observations provide more quantitative evidence that corroborates this conclusion (Supplementary Discussion 2).

Apparent loss from increased divergence and transverse heating was modeled via a Monte Carlo simulation (details in Methods section). Typical simulation results, shown in Fig. \ref{fig:slowinganddivergence}b, indicate that nearly all apparent molecule loss can be attributed to increased divergence and transverse heating. According to the simulation, the addition of transverse cooling to counteract divergence losses can vastly increase the flux of slow molecules. Other observations (e.g., dependence of the slowed LVP on laser power for various detunings) corroborate the basic picture that the observed slowed LVP is highly distorted by loss of slow molecules (Supplementary Discussion 3). Dependence of the slowing on various control parameters was generally weak over small ranges around their nominal values (Supplementary Discussion 4).

\begin{figure}
\includegraphics[height=1.35in]
{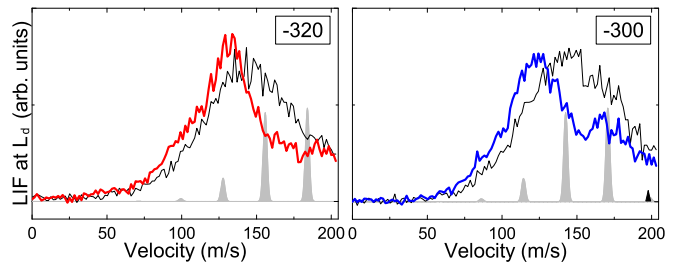}
\caption{{\bf Slowing with no applied magnetic field.} The $\Delta v_{00}^s$ detuning (in MHz) is shown in the upper right. Otherwise the representation is the same as in Fig. \ref{fig:slowinganddivergence}. Note the sharp features, in particular the increase in peak height or slope of the LVPs; these indicate longitudinal velocity compression within part of the distribution. This should be contrasted with the smooth LVPs obtained at large $B$ (Fig. \ref{fig:slowinganddivergence}). Here the $v_{00}^s$ modulation index is $m\!\!=\!\!2.6$.} \label{fig:sharpfeature}
\end{figure}

Most data was taken with $B\!\approx\!4\!-\!9$ G and $\theta_B\!=\!45^{\circ}$. Over this range, the slowed LVPs and LIF at  $L_s$ were fairly insensitive to the value of $B$. However, we observed that Earth's magnetic field, $B_E$, on its own allows some remixing of the dark Zeeman sublevels. Under certain conditions when $B=B_E$ (Supplementary Discussion 5), qualitatively different behavior was observed; namely, sharp features appeared in the LVP, as shown in Fig. \ref{fig:sharpfeature}. Moreover, under these conditions there is clear evidence for longitudinal velocity compression: the peak and slope of the LVP increases under these conditions for certain detunings $\Delta v_{00}^s$. We have been unable to find a simple explanation for these features, and full modeling of the system (including all $\sim\!\!33$ slowing laser frequencies, 44 molecular sublevels, $B$-field remixing, coherent dark states \cite{Berkeland2002}, etc.) is beyond the scope of this paper. However, this behavior could potentially be used to compress the molecular beam LVP. Ideally this would be done after slowing had already removed most of the kinetic energy from the beam, e.g., by using an initial region of large $B$ for broadband slowing, followed by a second region of small $B$ for longitudinal velocity compression and further slowing. A slow and nearly monoenergetic  beam would be ideal for trap loading.

In summary, we have demonstrated efficient radiation pressure slowing of an SrF molecular beam. Under certain conditions, we detect $\sim\!\!6\%$ of the initial detected flux at velocities $<\!\!50\frac{m}{s}$. The dominant loss mechanism at present is the increased divergence and transverse heating of the beam due to the slowing. The addition of transverse cooling should provide a much higher flux of slow molecules, suitable for loading a trap. It may be possible to use a low $B$-field section to compress the velocity distribution following the initial slowing. A slow molecular beam could be directly loaded into either a magneto-optical trap (MOT) \cite{Stuhl2008} or a sufficiently deep conservative trap, using optical pumping as a dissipative loading mechanism \cite{Riedel2011,RaymondOoi2003,Stuhler2001,DeMille2004}. Furthermore, the preliminary evidence of little loss during cycling, even after $\gtrsim\!\!10^4$ photons have been scattered, invites the possibility of moderately long lifetimes for SrF in a MOT.

{\small
\section{Methods}

We use an ablation-loaded cryogenic buffer gas beam source, which provides relatively low initial forward velocities, low internal temperatures, and high brightness \cite{Barry2011,Hutzler2011}. It produces an SrF molecular beam of $1.2 \!\times\! 10^{11}$ molecules/sr/pulse in the X$^2\Sigma^+_{1/2} (v \!\!=\!\!0,N\!\!=\!\!0)$ state. To mitigate variations of the molecular beam flux, data are taken by chopping the slowing lasers on and off between each ablation shot.  The background pressure in the beam propagation region is $\sim\!\!2\!\!\times\!\! 10^{-7}$ Torr.

With the $v_{00}^s$, $v_{10}^s$, and $v_{21}^s$ lasers applied, molecules cycle over the three bright ground states: X$(v\!\!=\!\!0,1,2; N\!\!=\!\!1)$. Due to different optical pumping rates into and out of these states, the $X(v\!\!=\!\!0,1; N\!\!=\!\!1)$ populations are expected to be comparable, while the X$(v\!\!=\!\!2,N\!\!=\!\!1)$ population should be significantly less. We hence employ a scheme to detect population in both X$(v\!\!=\!\!0,1; N\!\!=\!\!1)$ states. These states are excited to the A$(v\!\!=\!\!0,J\!\!=\!\!3/2)$ state via the $v_{00}^p$ and $v_{10}^p$ probe lasers, which are spatially overlapped and intersect the molecular beam at $L_d$. Both the $v_{00}^p$ and $v_{10}^p$ lasers have $f_{mod}\!\!=\!\!42$ MHz rf sidebands with $m\!\!=\!\!2.6$ to excite all SRS/HFS levels of the X$(v\!\!=\!\!0,1;N\!=\!1)$ states \cite{Shuman2009}; since they intersect the collimated molecular beam transversely, they are subject to negligible Doppler shift and broadening. The $v_{00}^p$ and $v_{10}^p$ laser powers are set to drive both transitions with the same Rabi frequency. Once excited to the D$(v\!\!=\!\!0,N\!\!=\!\!3,J\!\!=\!\!5/2)$ state by the longitudinally propagating $v_{AD}^p$ laser, the resulting LIF at $L_d$, predominantly at 360 nm, is filtered and measured by a photon-counting PMT. By scanning the $v_{AD}^p$ laser frequency and noting the Doppler shift (relative to the resonant frequency of a transverse $v_{AD}^p$ laser beam), we derive the molecular beam LVP. We verified that the detection efficiency and measured LVP are independent of whether the molecule is detected from X$(v\!\!=\!\!0,N\!\!=\!\!1)$ or X$(v\!\!=\!\!1,N\!\!=\!\!1)$. Power broadening from the $v_{AD}^p$ laser ($\sim\!\!23$ MHz FWHM) and magnetic field broadening ($\sim\!18$ MHz FWHM) lead to a measured broadening of $42$ MHz FWHM for the $v_{AD}^p$ detection profile, resulting in all velocity values having an uncertainty of $\pm 17\frac{m}{s}$.

The $v^s_{00}$, $v^s_{10}$, $v^s_{21}$, and $v^p_{AD}$ lasers are spatially overlapped using a combination of dichroic mirrors and a polarizing beamsplitter (PBS) to produce a single beam with $\frac{1}{e^2}$ intensity waist $d\!\!=\!\!3.4$ mm (except for the $v_{AD}^p$ laser with $d\!\!=\!\!4.4$ mm) counterpropagating to the molecular beam. The $v^s_{00}$, $v^s_{10}$, $v^s_{21}$ and $v^p_{AD}$ laser powers are 140 mW, 73 mW, 45 mW, and 70 mW respectively. The $v^s_{00}$, $v^s_{10}$, and $v^s_{21}$ laser detunings from resonance, denoted $\Delta v^s_{00}$, $\Delta v^s_{10}$, and $\Delta v^s_{21}$ respectively, are first varied iteratively to maximize LIF at $L_s$. For finer tuning, the $v^p_{AD}$ laser detuning, denoted $\Delta v_{AD}^p$, is set to resonantly excite SrF molecules with $v_{f}\!\approx\!50 \frac{m}{s}$, and $\Delta v^s_{00}$, $\Delta v^s_{10}$, and $\Delta v^s_{21}$ are varied iteratively to maximize the number of molecules detected in that Doppler class. Unless explicitly noted, $\Delta v^s_{10}$ and $\Delta v^s_{21}$ remain at these empirically determined values, denoted $\Delta v^{s,opt}_{10}$ and $\Delta v^{s,opt}_{21}$ respectively.

In the Monte Carlo simulation, particles are created at the source with randomized velocity distribution matching the measured forward and transverse velocity distributions of our source \cite{Barry2011}. We assume equal detection efficiency over the $v_{AD}^p$ $\frac{1}{e^2}$ beam waist at $L\!\!=\!\!L_d$. We estimate the force profile using a model 5-level system consisting of one excited state and four ground states (to match the four SRS/HFS levels). The degeneracy of the SRS/HFS levels and the accompanying level shifts and remixing within each SRS/HFS level due to the applied $B$-field are not included in the simulation. The saturation parameter $s$ is calculated for each of the four SRS/HFS levels assuming an estimated saturation power of 6 mW/cm$^2$ and the known $v_{00}^s$ laser rf sideband spectrum. Using $\tau$ and $s$, classical rate equations are solved to determine the equilibrium excited state population fraction, $\rho_{ee}$, as a function of the detuning from the center of the SRS/HFS spectrum, $\Delta$. The dependence of $\rho_{ee}$ is then fit to a Voigt profile. This process is repeated for the range of powers dictated by the $v_{00}^s$ laser's Gaussian intensity profile. Using the peak values of $\rho_{ee}$ for each intensity, we derive an estimate of how the maximum scattering rate varies with the distance from the center of the slowing beams, $r$. We finally model the scattering rate $R$ as the analytic function
\[
R(\Delta,r)\!=\!R_{max}\!\left[\!\frac{1}{1\!+\!(r/r_0)^a}\!\right] \!\left[\! N_1\int\limits_{-\infty}^{\infty}\!\frac{e^{-t^2/(2w_G^2)}}{(w_L/2)^2\!+\!(\Delta\!-\!t)^2}dt\!\right],
\]
where the normalization $N_1$ is chosen so that $R(0,0)\!\!=\!\! R_{max}$. The parameter values $r_0\! \!=\!\! 1.3$ mm, $a\!\!=\!\!3.75$, $w_L\!\!=\!\!99$ MHz and $w_G\!\!=\!\!95$ MHz are derived from these fits, without reference to the LVP data. The first two parameters control how the scattering rate varies with the beam intensity, while the latter two characterize the shape of the estimated scattering rate as a function of $\Delta$. Finally, the free parameter $R_{max}$ is varied manually to fit the LVP data for a variety of $\Delta$. We achieve good agreement with $R_{max}\!\!=\!\!2.8\times10^6$ s$^{-1}$, consistent with our previous observations \cite{Shuman2009}. Additional details of the simulation procedure are in Supplementary Methods.

}


\vspace{3 mm}

{\small
{\bf Acknowledgements} \hspace{3 mm} This material is based upon work supported by the ARO, the NSF and the AFOSR under the MURI award FA9550-09-1-0588.
\vspace{3 mm}

{\bf Author contributions} \hspace{3 mm} J.F.B. constructed the experiment, performed the measurements, and analyzed the data. E.S.S. assisted with each phase of the project. E.B.N. assisted with the experiment construction and data analysis. D.D. supervised each phase of the project.
\vspace{3 mm}

{\bf Competing financial interests} \hspace{3 mm} The authors declare that they have no competing financial interests
\vspace{3 mm}

{\bf Correspondence} \hspace{3 mm} Correspondence and requests for materials should be addressed to J.F.B. (email: john.barry@yale.edu).

\clearpage
\newpage

\section{Supplementary Material}

\subsection{Discussion 1}

The addition of transverse cooling shifts the simulated control LVP to lower velocities by $\lesssim \!10 \frac{m}{s}$ relative to the unperturbed simulated control LVP. This shift is due to the fact that, with transverse cooling, slower molecules are collimated closer to the source than are faster molecules, and the likelihood of their passing through the detection region is therefore greater. Overall, once the difference in scale has been accounted for, the shape of the simulated control LVP with transverse cooling is similar enough to the shape of the simulated control LVP without transverse cooling that the former is not shown in the main text. The simulated control LVP with and without transverse cooling is shown in Supplementary Fig. \ref{fig:controlwwotransversecooling} to illustrate this effect.

\subsection{Discussion 2}

We use the quantity $\Delta_{HM}$, defined as the shift of the half-maximum point on the leading edge of the observed slowed LVP (versus that of the control LVP), as one simple measure to evaluate the effectiveness of our slowing for different experimental parameters. Because slowed molecules are less likely to be detected (due to increased divergence, etc.), $\Delta_{HM}$ likely provides an underestimate of the actual slowing. With $\Delta v_{00}^s\!\!=\!\! -260$ MHz, providing resonant excitation for molecules with $v_{f}\!\!=\!\!175 \frac{m}{s}$, we routinely achieve $\Delta_{HM}\!\! \approx\!\!45\!-\!60 \frac{m}{s}$. Since the SrF recoil velocity is $v_{rc}\!\!=\!\!5.6 \frac{mm}{s}$, we interpret this as a mean number of photons scattered per molecule $\langle N_{sc}\rangle\!\!\approx\!\!10^4$, roughly an order of magnitude greater than in previous work \cite{Shuman2010}.

\subsection{Discussion 3}

Other behavior corroborates the basic picture that the observed slowed LVP is highly distorted by loss of slow molecules. As shown in Supplementary Fig. \ref{fig:hmsandlif}a, LIF at $L_s$ peaks at $\Delta v_{00}^s\!\!\approx \!\!-160$ MHz, whereas $\Delta_{HM}$ peaks at $\Delta v_{00}^s\!\!\approx\!\!-220$ MHz. These results are reproduced well by the simulation as shown in Supplementary Fig \ref{fig:hmsandlif}b. We interpret this discrepancy in peak locations to indicate that the greater scattering rate at $\Delta v_{00}^s=- 160 $ MHz is producing slower molecules, but these molecules are more likely than faster molecules to diverge before reaching the detection region. Further, consider the dependence of $\Delta_{HM}$ versus $v_{00}^s$ laser power, which is shown for $\Delta v_{00}^s=-340$ MHz, $-260$ MHz, and $-180$~MHz in Supplementary Fig. \ref{fig:powerdetuningbfield}. The data for each detuning are fit to a function of the form
\[
\Delta_{HM} = c_1\frac{P}{1+c_2 P}
\]
where $c_1$ and $c_2$ are fit constants and $P$ is the $v_{00}$ laser power. Note that $\Delta_{HM}$ saturates at lower powers when $\Delta v_{00}^s\!\! \approx\!\! \langle v_{f}\rangle/\lambda_{00}^s$, i.e. when the detuning most closely matches the Doppler shift of the molecular beam. We interpret this to suggest that more slowing is likely occurring for near-resonant detuning ($\sim\! -180$ MHz) than for further red detuning ($\sim\! -260$ MHz), but that at higher powers this additional slowing is not apparent in the data due to increased divergence and therefore decreased detection probability for the slowest molecules.

\begin{figure}
\renewcommand{\figurename}{Supplementary Figure}
\includegraphics[height=1.8in]
{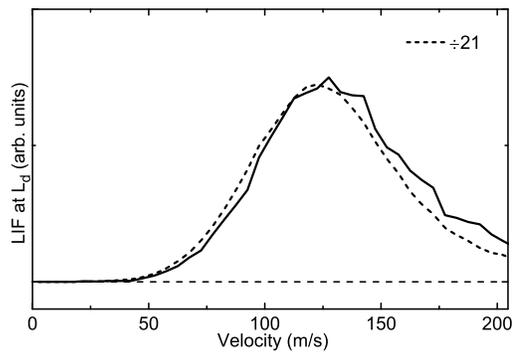} \caption{{\bf Impact of transverse cooling on simulated control longitudinal velocity profile.} The simulated control without transverse cooling (black solid line) and the simulated control with transverse cooling (black dashed line, scaled by 1/21) are very similar; the addition of transverse cooling shifts the simulated control LVP by $\lesssim 10 \frac{m}{s}$.} \label{fig:controlwwotransversecooling}
\end{figure}

\begin{figure}
\renewcommand{\figurename}{Supplementary Figure}
\includegraphics[height=2.2in]
{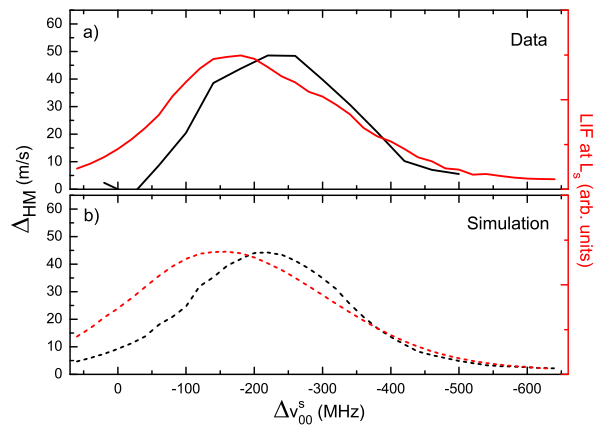} \caption{{\bf Dependencies of the half-maximum shift and spontaneous scattering fluorescence on laser detuning.} Measured (a) and simulated (b) dependence of $\Delta_{HM}$ and LIF at $L_s$ versus $\Delta v_{00}^s$. The LIF data is scaled to be the same height as the $\Delta_{HM}$ peak in both panels.} \label{fig:hmsandlif}
\end{figure}

\begin{figure}
\renewcommand{\figurename}{Supplementary Figure}
\includegraphics[height=2.2in]
{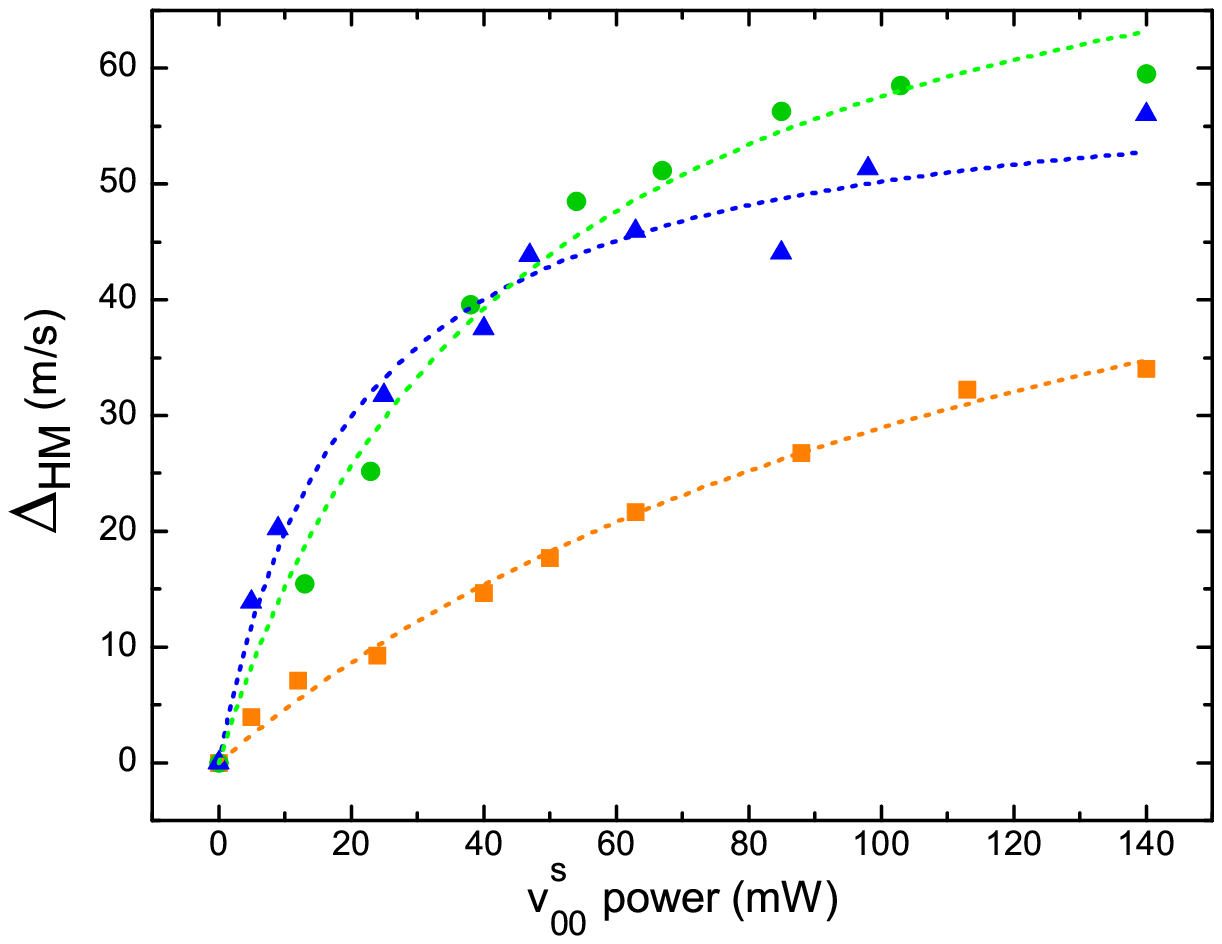} \caption{{\bf Dependencies of the half-maximum shift on laser power and detuning.}  $\Delta_{HM}$ versus $v_{00}^s$ laser power for $\Delta v_{00}^s$=
-340 MHz \small(\footnotesize${\color{Orange}\blacksquare}$\small),
-260 MHz \small(\large{\color{green}$\bullet$}\small),
-180 MHz ({\color{blue}$\blacktriangle$}) and associated fits.} \label{fig:powerdetuningbfield}
\end{figure}

\subsection{Discussion 4}

Overall, the slowed LVP was less sensitive to the $v_{10}^s$ and $v_{21}^s$ laser powers and detunings than to the $v_{00}^s$ laser power and detuning. At maximum $v_{10}^s$ power, $\Delta_{HM}$ was found to be insensitive when varying $\Delta v_{10}^s$ by $\sim\!\! 20$ MHz around $\Delta v_{10}^{s, opt}$. All $v_{21}^s$ powers $\gtrsim\!\!1$ mW produced indistinguishable slowed LVPs over an even broader range of detunings. Finally, under standard conditions where $B\!\!=\!\!4\!\!-\!\!9$ G, the slowed LVP was insensitive to changes in the $v_{00}^s$ sideband modulation index for $2.6\!<\!m\!<\!3.1$ and to changes in the $v_{00}^s$ sideband modulation frequency for $42$ MHz $<\! f_{mod}\! < 44$ MHz.

\subsection{Discussion 5}

In our apparatus, $B_{E} \! \approx \! .5$ G and the angle between $\vec{B}_{E}$ and the $v_{00}^s$ laser polarization is $\approx\!102^{\circ}$. The data shown in Fig. 4 of the main text were taken with a $v_{00}^s$ modulation index of $m\!\!=\!\!2.6$, where the sharp features were generally more pronounced than at the typical $m\!=\!3.1$ used for most of the slowing data.


\subsection{Supplementary Methods}

We found it necessary to adjust the size of the simulated source to a value larger than the physical cell exit hole diameter in order to obtain a qualitative match to our data. We believe this reflects the fact that collisions outside the source increase the effective area from which molecules are emitted. Our measurements show the molecular beam transverse velocity increases significantly ($\sim\!\! 2 \times$) within $\sim\!\! 10$ mm from the source \cite{Barry2011}, an effect we attribute to collisions with residual helium in the beam. Similar behavior was noted by a different group using an apparatus similar to ours \cite{Hutzler2011}. Given the measured initial transverse velocity, we calculate the spatial extent of the molecular beam at the distance where the transverse velocity stops increasing (indicating the beginning of molecular flow conditions). We find the calculated spatial extent of the molecular beam at this distance to be larger than 6.35 mm, the diameter of a hole in a collimating plate 34 mm downstream from the source. Ultimately the effective source diameter in the simulation was set to 6.35 mm; although the simulated source remained at the location of the physical source exit hole. Using a source diameter of 3 mm in the simulation (matching our physical cell exit hole) produced apparent losses greater than those measured by a factor of two or more for any set of reasonable parameters.


%
%
%

\end{document}